\pdfoutput=1
\documentclass[prapplied,aps,superscriptaddress,showpacs,reprint,longbibliography,nofootinbib]{revtex4-1}
\usepackage{lmodern}
\usepackage[T1]{fontenc}
\usepackage{amsmath,amssymb,amsfonts}
\usepackage{color}
\usepackage{graphicx}
\usepackage[normalem]{ulem}
\usepackage{hyperref}
\usepackage[utf8]{inputenc}
\usepackage{xcolor}

\usepackage[symbol]{footmisc}

\definecolor{LightBlue}{rgb}{0.55,0.83,0.78}

\definecolor{Green}{rgb}{0.1,0.9,0.5}

\hypersetup{
colorlinks=true,final=true,
        linkcolor=blue,
        citecolor=blue,
        filecolor=blue,
        filecolor=blue,
        urlcolor=blue,
        pdfauthor={S. Krinner and P. Kurpiers},
        pdftitle={Demonstration of an All-Microwave Controlled-Phase Gate between Far Detuned Qubits}
}

\begin{document}

\title{Demonstration of an All-Microwave Controlled-Phase Gate\\ between Far Detuned Qubits}

\author{S.~Krinner}
\email[]{skrinner@phys.ethz.ch}
\affiliation{Department of Physics, ETH Zurich, 8093 Zurich, Switzerland}

\author{P.~Kurpiers}
\thanks{S.K. and P.K. contributed equally to this work.}
\affiliation{Department of Physics, ETH Zurich, 8093 Zurich, Switzerland}

\author{B.~Royer}
\affiliation{Institut Quantique and D\'epartment de Physique, Universit\'e de Sherbrooke, Sherbrooke, Qu\'ebec J1K 2R1, Canada}
\affiliation{Department of Physics, Yale University, New Haven, Connecticut 06520, USA}
\author{P.~Magnard}
\affiliation{Department of Physics, ETH Zurich, 8093 Zurich, Switzerland}
\author{I.~Tsitsilin}
\affiliation{Department of Physics, ETH Zurich, 8093 Zurich, Switzerland}
\affiliation{Russian Quantum Center, National University of Science and Technology MISIS, Moscow 119049, Russia}
\author{J.-C.~Besse}
\affiliation{Department of Physics, ETH Zurich, 8093 Zurich, Switzerland}
\author{A.~Remm}
\affiliation{Department of Physics, ETH Zurich, 8093 Zurich, Switzerland}
\author{A.~Blais}
\affiliation{Institut Quantique and D\'epartment de Physique, Universit\'e de Sherbrooke, Sherbrooke, Qu\'ebec J1K 2R1, Canada}\affiliation{Canadian Institute for Advanced Research, Toronto, Canada}
\author{A.~Wallraff}
\email[]{andreas.wallraff@phys.ethz.ch}
\affiliation{Department of Physics, ETH Zurich, 8093 Zurich, Switzerland}

\date{\today}

\begin{abstract}
A challenge in building large-scale superconducting quantum processors is to find the right balance between coherence, qubit addressability, qubit-qubit coupling strength, circuit complexity and the number of required control lines. Leading all-microwave approaches for coupling two qubits require comparatively few control lines and benefit from high coherence but suffer from frequency crowding and limited addressability in multi-qubit settings. Here, we overcome these limitations by realizing an all-microwave controlled-phase gate between two transversely coupled transmon qubits which are far detuned compared to the qubit anharmonicity. The gate is activated by applying a single, strong microwave tone to one of the qubits, inducing a coupling between the two-qubit $|f,g\rangle$ and $|g,e\rangle$ states, with $|g\rangle$, $|e\rangle$, and $|f\rangle$ denoting the lowest energy states of a transmon qubit. Interleaved randomized benchmarking yields a gate fidelity of $97.5\pm 0.3 \%$ at a gate duration of $126\,\rm{ns}$, with the dominant error source being decoherence. We model the gate in presence of the strong drive field using Floquet theory and find good agreement with our data. Our gate constitutes a promising alternative to present two-qubit gates and could have hardware scaling advantages in large-scale quantum processors as it neither requires additional drive lines nor tunable couplers.

\end{abstract}

\maketitle

\section{Introduction}

Superconducting circuits making use of the concepts of circuit quantum electrodynamics \cite{Blais2020a} constitute a promising platform for quantum computing. Recently, processors containing several tens of superconducting qubits have been demonstrated \cite{Cross2018, Otterbach2017, Arute2019}. While high-fidelity single-qubit operations with error rates below 0.1\% are routinely achieved, two-qubit gate errors are typically at the percent level \cite{Kjaergaard2020a, Gambetta2017}, with only a few recent experiments achieving two-qubit gate errors of a few per mill \cite{Barends2019, Foxen2020, Kjaergaard2020}. Hence, two-qubit gates limit the performance of state-of-the-art quantum processors and a variety two-qubit gate schemes are currently explored. One typically distinguishes between two classes of approaches, flux-activated and microwave-activated gates.


The first class relies on the dynamic flux tunability of either the qubits or a separate coupling circuit. In this class gates are activated by tuning the qubits in frequency to fulfill certain resonance conditions between two-qubit states \cite{DiCarlo2009, Barends2014, Chen2014m, McKay2015, Rol2019} or by parametrically modulating the qubit transition frequency \cite{McKay2016, Caldwell2018, Mundada2019}. The main benefit are short gate times, which however come at the cost of degraded coherence times or crossings with two-level-system defects \cite{Klimov2018} when tuning the qubit frequency away from its so-called sweet spot frequency, at which the qubit is first-order insensitive to flux noise \cite{Koch2007}.

In the second class of approaches the qubits are fixed in frequency and two-qubit interactions are activated using a microwave tone \cite{Leek2009, Chow2012, Poletto2012, Chow2013, Cross2015, Egger2019}.
The main advantage of this approach is its potentially higher coherence when using fixed frequency qubits or frequency-tunable qubits operated at their flux sweet spot. In addition, control electronics and wiring requirements are somewhat lower as no flux control lines are needed. Instead, one resorts to the same control and pulse shaping hardware as also used for the realization of single-qubit gates.
The main disadvantage of all-microwave approaches is the typically longer gate time \cite{Leek2009, Chow2012, Poletto2012, Chow2013, Cross2015, Egger2019}.

The cross-resonance gate \cite{Rigetti2010, Chow2012, Sheldon2016}, in particular, constitutes one of the most frequently used all-microwave gates. However, for this gate to work, the detuning between the two qubits has to be smaller than the anharmonicity of the qubits. For multi-qubit devices this condition imposes stringent requirements on fabrication precision of Josephson junctions and leads to frequency crowding \cite{Brink2018}, eventually reducing gate speed and qubit addressability due to a higher sensitivity to cross talk. Here, we present an all-microwave controlled-phase gate which allows for large detunings compared to the anharmonicity. Our gate is simple and resource-friendly as it requires only a single microwave drive tone applied to one of the qubits in contrast to two drive tones \cite{Leek2009, Sheldon2016, Egger2019}, does not require re-focusing pulses during the gate \cite{Corcoles2013}, nor does it make use of real photons in an additional resonator \cite{Leek2009, Egger2019, Cross2015}. The only requirements are a transverse coupling between the qubits and a strong microwave drive.




\section{System and Setup}

The Hamiltonian describing two transversally coupled transmon qubits A and B in presence of a drive on qubit A reads
\begin{align}
\hat{H}/\hbar =& \sum\limits_{i=A,B} \omega_i \hat{a}_i^{\dagger} \hat{a}_i + \frac{\alpha_i}{2} \hat{a}_i^{\dagger} \hat{a}_i^{\dagger} \hat{a}_i \hat{a}_i \nonumber\\
& + J\left(\hat{a}_{\rm{A}}^{\dagger}\hat{a}_{\rm{B}} + \hat{a}_{\rm{A}}\hat{a}_{\rm{B}}^{\dagger}\right) + \Omega_{\rm{A}}(t)\left(\hat{a}_{\rm{A}}^{\dagger} + \hat{a}_{\rm{A}} \right),\label{eqn:Ham}
\end{align}
with $\hat{a}_i$ ($\hat{a}_i^{\dagger}$) the lowering (raising) operator of qubit $i$, and $\Omega_{\rm{A}}(t)$ a microwave drive applied to qubit A.

\begin{figure}
     \center
     \includegraphics{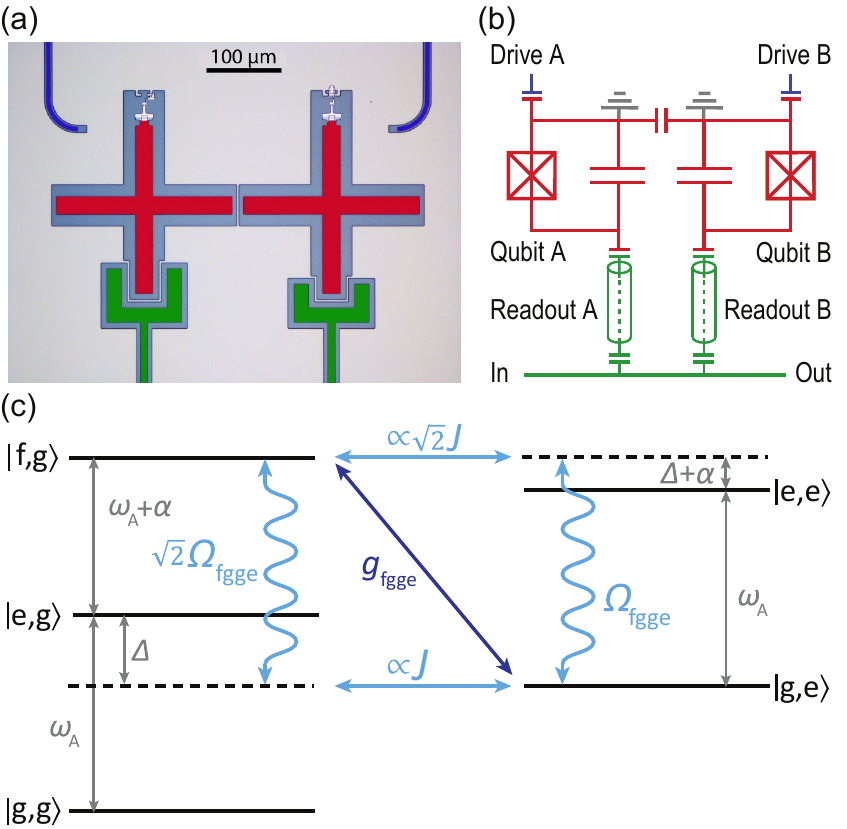}
     \caption{{Device and gate scheme.} (a) Micrograph and (b) circuit diagram of the core elements of the sample. Each of the two capacitively coupled transmon qubits (red) is coupled to an individual readout resonator (green) and drive line (blue). (c) Schematic of the energy-level diagram of qubit A with qubit B in the $|g\rangle$ state (left) and in the $|e\rangle$ state (right). Virtual states are indicated by dashed lines.}
     \label{fig1}
\end{figure}

The superconducting device used in our experiment uses a frequency-tunable transmon qubit (qubit A) and a fixed-frequency transmon qubit (qubit B). The first qubit is made tunable to provide more freedom in the choice of operation frequencies, but could be at fixed frequency as well.
The two qubits have frequencies $\omega_{\rm{A}}/2\pi=6.496\,$GHz and $\omega_{\rm{B}}/2\pi=4.996\,$GHz, energy relaxation times $T_{\rm{1,A}}=7\,\mu$s and $T_{\rm{1,B}}=20\,\mu$s, anharmonicities $\alpha_{\rm{A}}/2\pi = -257\,$MHz and $\alpha_{\rm{B}}/2\pi = -271\,$MHz, and are capacitively coupled with a coupling strength $J/2\pi=42(1)\,$MHz, see Fig.~\ref{fig1}(a) and (b). We control the state of each qubit using amplitude and phase modulated microwave pulses \cite{Motzoi2009, Chow2010, Gambetta2011a}, which are generated by upconverting the signals from an arbitrary waveform generator and applied to the qubits through a dedicated drive line. Prior to each experimental run we reset the qubits using the protocol introduced in \cite{Magnard2018}, reducing the excited state populations of qubit A and B to $0.6\%$ and $0.8\%$, respectively (see Appendix~\ref{app:Parameters} for details).

For qubit readout, two resonators at frequencies $\omega_{\rm{r,A}}/2\pi=7.379\,$GHz and $\omega_{\rm{r,B}}/2\pi=7.076\,$GHz are dispersively coupled to qubit A and qubit B, respectively, with strength $g_{\rm{A}}/2\pi=52\,$MHz and $g_{\rm{B}}/2\pi=71\,$MHz. Both resonators are coupled to a common feedline with coupling rates $\kappa_{\rm{A}}/2\pi=0.67\,$MHz and $\kappa_{\rm{B}}/2\pi=0.63\,$MHz. We determine the $|g\rangle$, $|e\rangle$, and $|f\rangle$ state population of both qubits by applying two gated microwave tones to the feedline of the readout resonators at frequencies and powers optimized for qutrit readout \cite{Bianchetti2010}. The transmitted signal is amplified at $10 \,\rm{mK}$ by a traveling wave parametric amplifier \cite{Macklin2015} and at 4\,K by a high-electron-mobility transistor amplifier. At room temperature the signal is further amplified, split into two paths, which are separately down-converted using an I-Q mixer, digitized using an analog-to-digital converter, digitally down-converted and processed using a field programmable gate array \cite{Salathe2018}. We extract the qutrit populations of each transmon using single-shot readout. We record each measurement trace 2000 (4000) times for all characterization (randomized benchmarking) experiments and account for readout errors \cite{Dewes2012, Kurpiers2018} (see Appendix~\ref{app:Parameters}).


\section{Gate Concept}

Our gate exploits a Raman transition between the two-qubit states $|f, g\rangle$ and $|g, e\rangle$. The transition is analogous to the cavity-assisted Raman transition used recently for photon shaping and remote quantum communication \cite{Pechal2014, Zeytinoglu2015, Kurpiers2018, Gasparinetti2016}, qutrit reset \cite{Magnard2018, Egger2018} and two-qubit gates \cite{Egger2019}, with the distinction that here the cavity is replaced by a second qubit. The coupling between $|f, g\rangle$ and $|g, e\rangle$ is activated by a strong microwave tone $\Omega_{\rm{A}}(t) = \Omega_{\rm{fgge}} \cos(\omega_{\rm{fgge}} t)$ applied to the drive line of qubit A at a frequency corresponding to the energy difference between the two states, i.~e.~at $\omega_{\rm{fgge},0}/2\pi=(\omega_{\rm{A}}+\Delta+\alpha_{\rm{A}})/2\pi \approx 7.739\,$GHz, with $\Delta=\omega_{\rm{A}}-\omega_{\rm{B}}$ and the subscript '0' labeling the unshifted transition frequency in absence of a drive-induced ac-Stark shift on qubit $A$. 
The coupling is mediated by virtual states, which are coupled to $|f, g\rangle$ and $|g, e\rangle$ via the drive $\Omega_{\rm{fgge}}$ and the direct qubit-qubit coupling $J$, see Fig.~\ref{fig1}(c).
The two coupling paths between $|f, g\rangle$ and $|g, e\rangle$ indicated by the light blue arrows interfere destructively and give rise to a total coupling strength of
\begin{equation}
\label{eq:fggeLinearCoupling}
g_{\rm{fgge}}=\frac{\Omega_{\rm{fgge}} J \alpha_{\rm{A}}}{\sqrt{2}\Delta(\Delta+\alpha_{\rm{A}})}.
\end{equation}

\noindent
Due to the large detuning between the qubits, a large drive amplitude is required to reach a coupling strength of a few MHz and thus a gate time $1/(2\,g_{\rm fgge})$ significantly below 1\,$\mu$s.

When driving the fg-ge transition for a duration which corresponds to a full round trip in the fg-ge manifold the state $|g,e\rangle$ picks up a geometric phase of $\pi$ \cite{Sjoqvist2015}, 
thereby realizing a controlled-phase gate. Using virtual-Z gates \cite{McKay2017}, this conditional phase can be assigned to either of the computational states. We perform a virtual-Z gate on qubit B, so that the state $|e,e\rangle$ effectively picks up the phase, corresponding to flux-based implementations of controlled-phase gates which exploit the coupling between the $|e,e\rangle$ and the $|g,f\rangle$ state \cite{DiCarlo2009,Barends2014,Caldwell2018}.

\section{Gate Calibration}

The fg-ge pulse is realized as a flat-top envelope with Gaussian rising and falling edges with widths $\sigma=5\,$ns truncated at $3\,\sigma$, carrier frequency $\omega_{\rm{fgge}}/2\pi$, normalized amplitude $A_{\rm{fgge}}$, and duration $\tau_{\rm{fgge}}$.
%
\begin{figure}[tb]
     \center
     \includegraphics{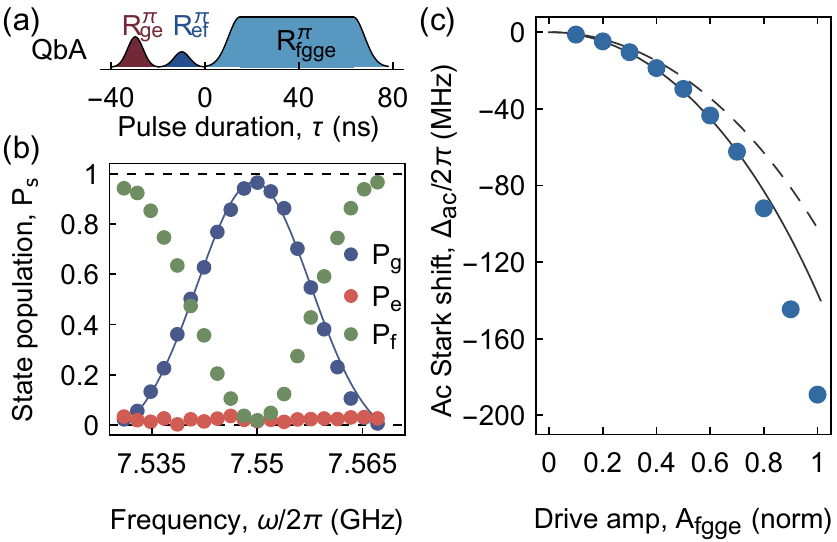}
     \caption{{Calibration of ac-Stark shift.} (a) Pulse sequence applied to qubit A to resolve the fg-ge transition in a pulsed spectroscopy experiment. $\rm{R}^{\pi}_{\rm{ge}}$ and $\rm{R}^{\pi}_{\rm{ef}}$ label Gaussian derivative-removal-by-adiabatic-gate (DRAG) microwave pulses~\cite{Motzoi2009,Chow2010,Gambetta2011a} for the transmon transitions $g\leftrightarrow e$ and $e\leftrightarrow f$ of angle $\pi$. $\rm{R}^{\pi}_{\rm{fgge}}$ labels a flattop pulse on the transmon-transmon transition $fg\leftrightarrow ge$. (b) Qutrit Populations $\rm{P}_{\rm{g,e,f}}$ of qubit A versus the frequency $\omega/2\pi$ of a flattop fg-ge pulse with amplitude $\rm{A}_{\rm{fgge}}=1.0$. The solid line is a Gaussian fit from whose center we extract the ac-Stark shift. (c) Measured ac-Stark shifts $\Delta_{\rm{ac}}$ of the fg-ge transition (blue dots) versus drive amplitude $A_{\rm{fgge}}$. The solid line is calculated from numerical simulations based on Floquet theory, while the dashed line results from simulations based on Eq.~(\ref{eqn:Ham}) in the rotating-wave approximation.}
     \label{fig2}
\end{figure}

Due to the ac-Stark effect the fg-ge transition frequency $\omega_{\rm{fgge}}$ depends on the drive amplitude $\Omega_{\rm{fgge}}$. Similar to \cite{Magnard2018}, we calibrate the ac-Stark shift by preparing the qubits in the $|f,g\rangle$ state, applying the fg-ge pulse, and reading out the state of qubit A, see Fig.~\ref{fig2}(a). For a given $A_{\rm{fgge}}$ we adjust $\tau_{\rm{fgge}}$ to obtain Rabi angles close to $\pi$ and measure the $|g\rangle$ state population of qubit A as a function of frequency, see Fig.~\ref{fig2}(b). On resonance, the population transfer from $|f,g\rangle$ to $|g,e\rangle$ is maximum. We fit the resulting spectrum to a Gaussian from whose center $\omega_{\rm{fgge}}$ we infer the ac-Stark shift $\Delta_{\rm{ac}}=\omega_{\rm{fgge}}-\omega_{\rm{fgge}, 0}$ of the fg-ge transition frequency. In this way we measure the dependence of $\Delta_{\rm{ac}}$ on $A_{\rm{fgge}}$, see Fig.~\ref{fig2}(c). Due to the large drive amplitude, we observe deviations from a quadratic dependence \cite{Magnard2018}, as discussed below.

\begin{figure}[tb]
     \center
     \includegraphics{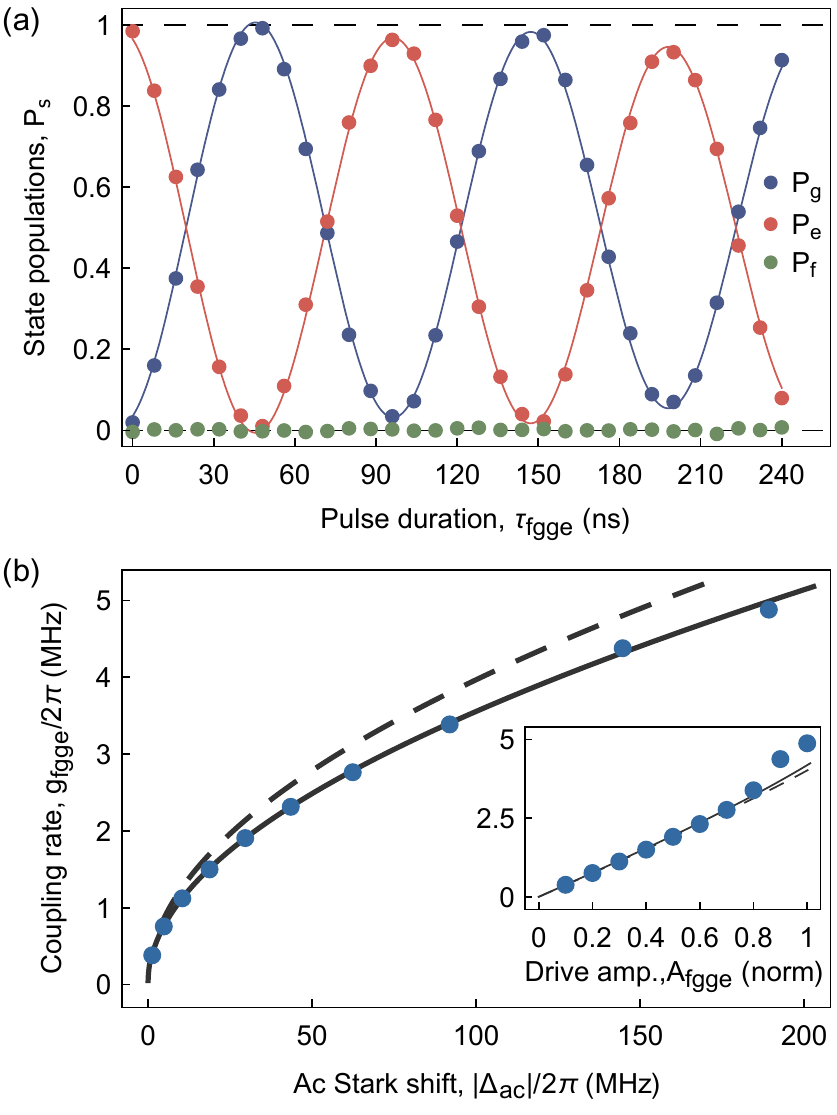}
     \caption{{Rabi oscillations.} (a)~Qutrit populations $\rm{P}_{\rm{g,e,f}}$ of qubit B versus pulse duration $\tau_{\rm{fgge}}$ of a resonant flat-top fg-ge pulse with amplitude $\rm{A}_{\rm{fgge}}=1.0$ corresponding to Rabi oscillations between $|f,g\rangle$ and $|g,e\rangle$. The initially prepared state is $|ge\rangle$. Solid lines are exponentially decaying sinusoidal fits. (b)~Extracted coupling strength $g_{\rm fgge}$ {\sl vs.}~drive-induced ac-Stark shift $\Delta_{\rm{ac}}$. Blue dots are experimental data. The solid line is calculated from numerical simulations based on Floquet theory, while the dashed line results from simulations for which a rotating-wave approximation has been applied to Eq.~(\ref{eqn:Ham}). Inset: Coupling strength $g_{\rm fgge}$ {\sl vs.}~drive amplitude $A_{\rm fgge}$.} 
     \label{fig3}
\end{figure}

We next measure the coupling strength $g_{\rm{fgge}}$ {\sl vs.}~$A_{\rm{fgge}}$ in a Rabi experiment. For a given $A_{\rm{fgge}}$, we prepare $|g, e\rangle$, apply the fg-ge pulse at the previously determined resonance frequency for variable $\tau_{\rm{fgge}}$ and  measure the qutrit populations of qubit B. We fit the resulting Rabi oscillations with an exponentially decaying sinusoidal function, see Fig.~\ref{fig3}(a) for an example with $A_{\rm{fgge}}=1.0$. The coupling strength $g_{\rm{fgge}}$ is given by half of the fitted Rabi oscillation frequencies. For the largest drive amplitude $A_{\rm{fgge}}=1.0$ we achieve a coupling strength $g_{\rm{fgge}}/2\pi=5.0\,$MHz. We plot the extracted $g_{\rm{fgge}}/2\pi$ as a function of $\Delta_{\rm ac}$ [Fig.~\ref{fig3}(b)] rather than the voltage amplitude $A_{\rm fgge}$ set at the instrument in order to be insensitive to possible nonlinearities between $A_{\rm fgge}$ and the drive amplitude $\Omega_{\rm fgge}$ at qubit A, see also Appendix~\ref{app:calibration}.
%

We obtain very good agreement between data and a numerical model based on Floquet theory with independently determined parameters [solid line in Fig.~\ref{fig3}(b)]. The model takes into account counter-rotating terms induced by the drive and the full cosine potential of the transmon qubits, see Appendix~\ref{app:calibration} for details. For comparison, simulations based on a rotating-wave approximation to Hamiltonian Eq.~(\ref{eqn:Ham}) fail to accurately describe our data [dashed line in Fig.~\ref{fig3}(b)]. Due to the large drive amplitude (for $A_{\rm fgge}=1$ we estimate $\Omega_{\rm fgge}/\omega_{\rm fgge}\sim 0.15\,$) counter-rotating terms in the Hamiltonian are important.
For completeness, we also plot $g_{\rm{fgge}}$ {\sl vs.}~$A_{\rm fgge}$ [Fig.~\ref{fig3}(b) inset]. For this data as well as for the data $\Delta_{\rm ac}$ {\sl vs.}~$A_{\rm fgge}$ presented in Fig.~\ref{fig2}(c) we observe deviations from theory for $A_{\rm fgge}>0.7$, which we attribute to a non-linearity between $A_{\rm fgge}$ and the effective drive amplitude $\Omega_{\rm fgge}$ at qubit A, see Appendix~\ref{app:calibration}.

\begin{figure}[tb]
     \center
     \includegraphics{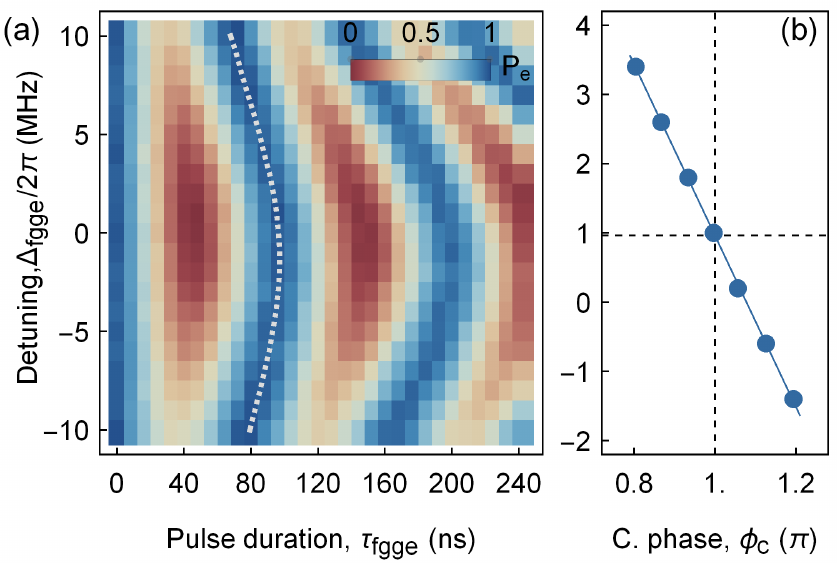}
     \caption{
      (a)~Excited state population $\rm{P}_{\rm{e}}$ of qubit B as a function of pulse duration $\tau_{\rm{fgge}}$ and detuning $\Delta_{\rm{fgge}}$ of the fg-ge drive tone. The dashed line indicates the extracted times for which the population recovery in the computational subspace is maximum. (b)~Conditional phase $\phi_c$ versus detuning $\Delta_{\rm{fgge}}$ extracted from Ramsey experiments. Horizontal and vertical dashed lines indicate the parameters for which $\phi_c=\pi$.}
     \label{fig4}
\end{figure}

To implement a controlled-phase gate it is important to take into account the dispersive always-on coupling of the qubits. In the dispersive approximation $\Delta\gg J$ the exchange coupling term in Eq.~(\ref{eqn:Ham}) transforms into $\chi\hat{a}_{\rm{A}}^{\dagger}\hat{a}_{\rm{A}} \hat{a}_{\rm{B}}^{\dagger}\hat{a}_{\rm{B}}$, with $\chi = 2 J^2(\alpha_{\rm{A}}+\alpha_{\rm{B}})/[(\Delta+\alpha_{\rm{A}})(\Delta-\alpha_{\rm{B}})]$ \cite{Barends2014}. From a Ramsey experiment we determine $\chi/2\pi=-0.83(1)\,$MHz, in agreement with the calculated value of -0.85(4)\,MHz and comparable to the values found in Ref.~\cite{Barends2014}. Hence, the $|e,e\rangle$ state acquires not only a conditional geometric phase $\phi_{\rm{fgge}}$  due to the rotation in the $|fg\rangle$-$|ge\rangle$ subspace (assuming a virtual-Z gate on qubit B), but also a conditional dynamical phase $\phi_{\rm{zz}}$ due to the disperse always-on coupling of the qubits. As a result, $\phi_{\rm{fgge}}$ has to be smaller than $\pi$. Under the constraint of full population recovery into the computational subspace, this is achieved by driving the fg-ge transition slightly off-resonantly at a frequency $\omega_{\rm{fgge}}+\Delta_{\rm{fgge}}$, with $\Delta_{\rm{fgge}}$ the detuning between the drive and the fg-ge transition frequency, see Appendix~\ref{app:CphaseCalibration}.
We measure the corresponding Rabi oscillations as a function of $\Delta_{\rm{fgge}}$ for $A_{\rm{fgge}}=1.0$ and obtain the characteristic Chevron-like pattern shown in Fig.~\ref{fig4}(a).

While the dispersive coupling can be taken into account in the calibration of the gate, we note that it leads to coherent errors in multi-qubit settings \cite{Krinner2020}. Possible mitigation strategies without compromising gate time include reducing the transversal coupling strength while increasing the drive strength, making use of dynamical decoupling techniques \cite{Viola1998,Vandersypen2004,Bylander2011,Guo2018a}, combining qubits with opposite anharmonicity since $\chi\propto\alpha_{\rm{A}}+\alpha_{\rm{B}}$ \cite{Ku2020}, and driving the fg-ge transition off-resonantly during idle times, which allows for adjusting and canceling the dispersive interaction \cite{Rosenblum2018b}.

To calibrate the controlled-phase gate we follow a two-step procedure. First, we measure the conditional phase $\phi_c = \phi_{\rm{fgge}} + \phi_{\rm{zz}}$ as a function of $\Delta_{\rm{fgge}}$. For this purpose, we extract the pulse durations $\tau_{\rm{fgge}}$ for which qubit B is back in the $|e\rangle$ state, see dashed line in Fig.~\ref{fig4}(a). This condition corresponds to minimum $|f\rangle$ level population of qubit A and therefore to maximum population recovery into the computational subspace. We then measure $\phi_c$ using a Ramsey experiment on qubit B while driving the fg-ge transition on qubit A, which is prepared in either $|g\rangle$ or $|e\rangle$. The difference between the phases extracted from both measurements yields $\phi_c$, see Fig.~\ref{fig4}(b). From a linear fit to the data we extract the detuning $\Delta_{\rm{fgge},\pi}/2\pi= 0.962\,$MHz which yields $\phi_c=(1.00\pm0.01)\pi$. The second step consists of calibrating the single-qubit phases $\phi_{s,i}$ with $i=A,B$, which are affected by the fg-ge drive induced ac-Stark effect. We measure $\phi_{s,i}$ using a Ramsey experiment on qubit $i$ in the presence of the fg-ge pulse on qubit A and correct these phases using
virtual-Z gates.


\section{Gate Characterization}

\begin{figure}[tb]
     \center
     \includegraphics{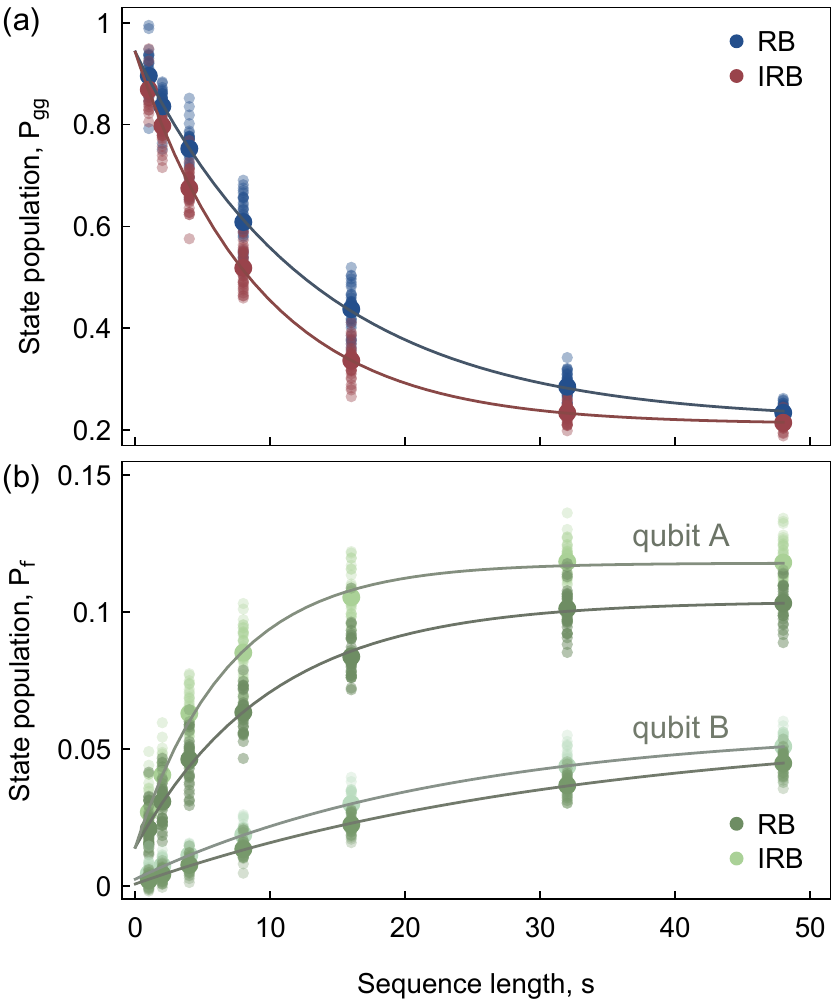}
     \caption{{Randomized benchmarking.} (a)~Population $\rm{P}_{\rm{gg}}$ and (b)~$\rm{P}_{\rm{f}}$ versus two-qubit Clifford group gate sequence length $s$. The number of randomly generated sequences is 36. Solid lines in (a) are exponential fits used to extract the gate fidelity. Solid lines in (b) are fits to a rate equation model used to extract the leakage error per gate.}
     \label{fig5}
\end{figure}

We finally characterize the gate by performing interleaved randomized benchmarking \cite{Gaebler2012, Magesan2012}. We obtain a controlled-phase gate fidelity of 97.5(3)\%, extracted from exponential fits of the form $A+B p^s$ to the interleaved measurement and to a reference measurement, see red and blue data points in Fig.~\ref{fig5}(a) respectively. Here, $p$ denotes the depolarizing parameter, $s$ is the number of applied two-qubit Clifford gates, and $A$, $B$ are coefficients accounting for state preparation and measurement (SPAM) errors \cite{Magesan2012}. The fidelity of the reference measurement is 94.5(1)\%.

Our qutrit readout allows us to simultaneously extract the leakage rate to the $|f\rangle$ level of both qubits. We fit the observed rise in $|f\rangle$ level population $P_{\rm f}$ [Fig.~\ref{fig5}(b)] as a function of sequence length $s$ to a rate equation model \cite{Chen2016} of the form $P_{\rm f}(s)=p_{\infty}(1-{\rm e}^{-\Gamma s})+p_{0} {\rm e}^{-\Gamma s}$, with $p_{0}$ the initial $|f\rangle$ level population, $p_{\infty}=\gamma_{\uparrow}/\Gamma$ the asymptotic $|f\rangle$ level population, and $\Gamma=\gamma_{\uparrow}+\gamma_{\downarrow}$ the sum of the leakage rate $\gamma_{\uparrow}$ and the decay rate $\gamma_{\downarrow}$. Subtracting the reference leakage rate $\gamma_{\uparrow,{\rm RB}}$ from the leakage rate of the interleaved experiment, $\gamma_{\uparrow,{\rm IRB}}$, we extract leakage errors per controlled-phase gate of $\epsilon_{\rm{l,A}}=0.7(3)\%$ and $\epsilon_{\rm{l,B}}=0.07(2)\%$ for qubit A and B, respectively. As expected, the leakage error for qubit A is significantly larger than for qubit B because only the $|f\rangle$ level of qubit A is populated during the gate.

From master equation simulations we compute an average gate fidelity of 97.5\%, which is in good agreement with the measured fidelity and indicates that the gate fidelity is limited by decoherence. The numerical simulation reveal that 0.4\% leakage per gate can be attributed to $T_{2ef}$ errors on qubit A, while the remaining leakage errors are caused by other decoherence channels. In particular, due to the dressing of the states in the driven basis, different decoherence channels can contribute. Removing transmon decoherence from master equation simulations we estimate that a gate fidelity higher than 99.9\% is possible without pulse optimization.

\section{Summary}

In summary, we have demonstrated a fast, coherence limited all-microwave controlled-phase gate between two qubits which are detuned by about six times the qubit anharmonicity. In particular, the gate imposes no constraints on the qubit-qubit detuning and is activated by a single microwave tone applied to the drive line of one of the qubits. Hence, no further resources beyond those already used for single-qubit gates are required. We therefore believe that in future multi-qubit quantum processors our gate will provide hardware scaling advantages compared to processors relying on fast flux tunability of qubits \cite{Andersen2020b} and tunable coupling circuits \cite{Arute2019}. This assumes that the relatively large always-on dispersive coupling can be mitigated without large overhead \cite{Viola1998,Vandersypen2004,Bylander2011,Guo2018a,Ku2020,Rosenblum2018b}.
Finally, the engineered coupling between $|f,g\rangle$ and $|g,e\rangle$ can be used in a heralded quantum communication protocol \cite{Kurpiers2019}, where an auxiliary qubit indicates photon loss events.
\section*{Acknowledgements}
We thank A. Akin for programming the FPGA firmware, M. Collodo for contributions to the measurement setup, C. K. Andersen and C. Eichler for discussion, and C. Le Calonnec and A. Petrescu for help with the Floquet simulations. This work was supported by the European Research Council (ERC) through the 'Superconducting Quantum Networks' (SuperQuNet) project, by the National Centre of Competence in Research 'Quantum Science and Technology' (NCCR QSIT) a research instrument of the Swiss National Science Foundation (SNSF), by the Office of the Director of National Intelligence (ODNI), Intelligence  Advanced  Research  Projects  Activity  (IARPA), via the U.S. Army Research Office grant W911NF-16-1-0071, by the SNFS R’equip grant 206021-170731, by ETH Zürich, by NSERC, the Canada First Research Excellence Fund, and by the Vanier Canada Graduate Scholarships. S. Krinner acknowledges financial support by Fondation Jean-Jacques \& Felicia Lopez-Loreta and the ETH Zurich Foundation.
I. Tsitsilin acknowledges partial support from the Ministry of Education and Science of the Russian Federation in the framework of Increase Competitiveness Program of the National University of Science and Technology MISIS (Contract No. K2-2017-081).
The views and conclusions contained herein are those of the authors and should not be interpreted as necessarily representing the official policies or endorsements,either expressed or implied, of the ODNI, IARPA, or the U.S. Government.
\appendix
\section{Sample and Setup}
\label{app:Parameters}

\begin{figure}[tb]
     \center
     \includegraphics[width=8.6cm]{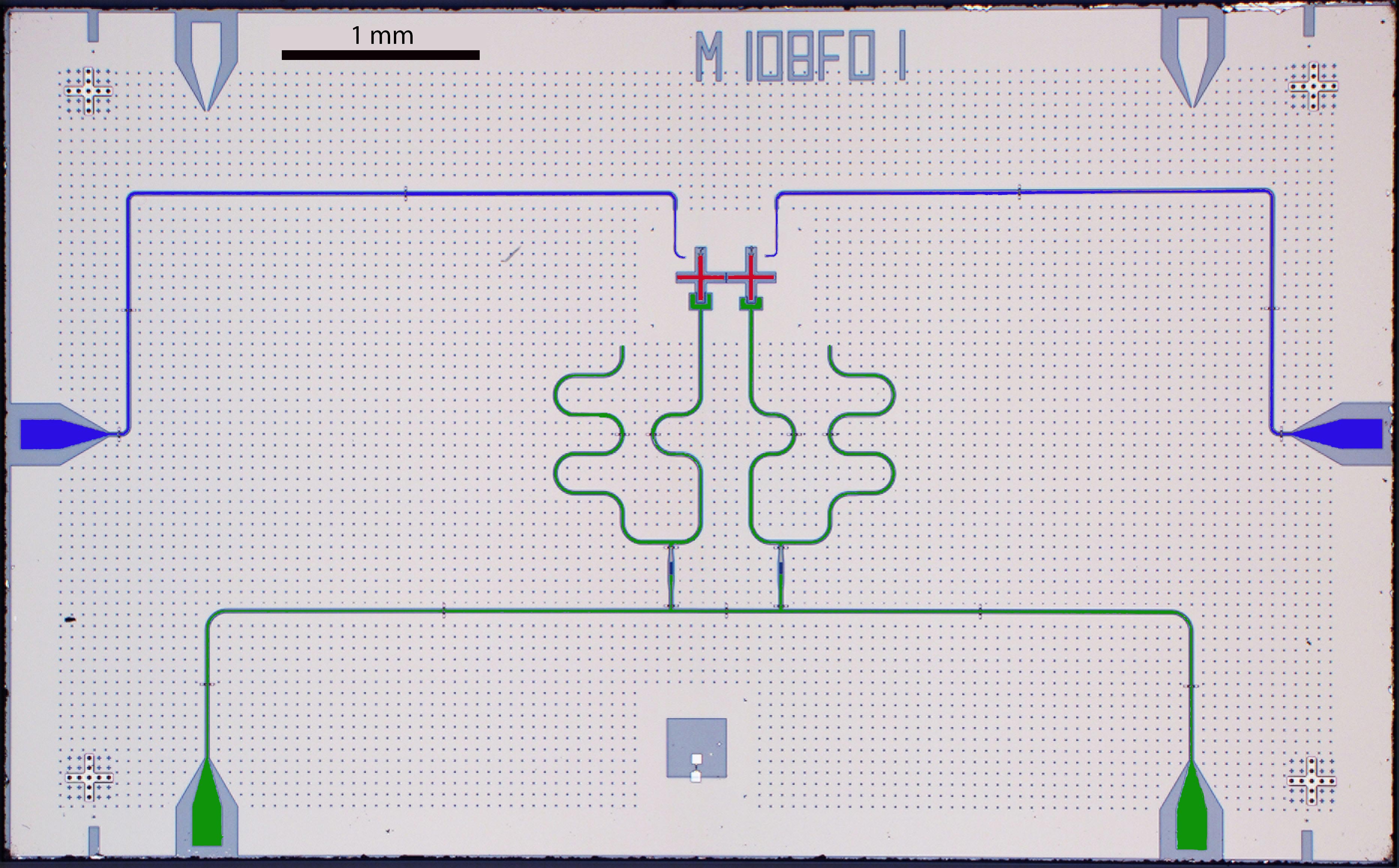}
     \caption{False-colored micrograph of the superconducting device showing the transmon qubits in red, the readout circuit in green, and the qubit drive lines in blue.}
     \label{fig6}
\end{figure}

The superconducting device is made of a patterned Niobium thin film on a high-resistivity Silicon substrate using standard photolithography techniques, see Fig.~\ref{fig6}. Josephson junctions are fabricated using electron beam lithography and shadow evaporation of aluminium with lift-off. Qubit drive and fg-ge drive signals are combined before being amplified at room temperature and routed to the dilution refrigerator. We use either single side-band modulation with IQ-mixers driven by a local oscillator (LO) and an arbitrary waveform generator (AWG), or alternatively, directly synthesized drive-pulses from a high-bandwidth AWG (fAWG).

The input lines are thermalized at each temperature stage of the dilution refrigerator and are attenuated at the 4K-, CP- and BT-stages \cite{Krinner2019}. We use a superconducting coil to thread flux through the superconducting quantum-interference device (SQUID) of qubit A to tune its frequency.

\begin{figure}[tb]
     \center
     \includegraphics{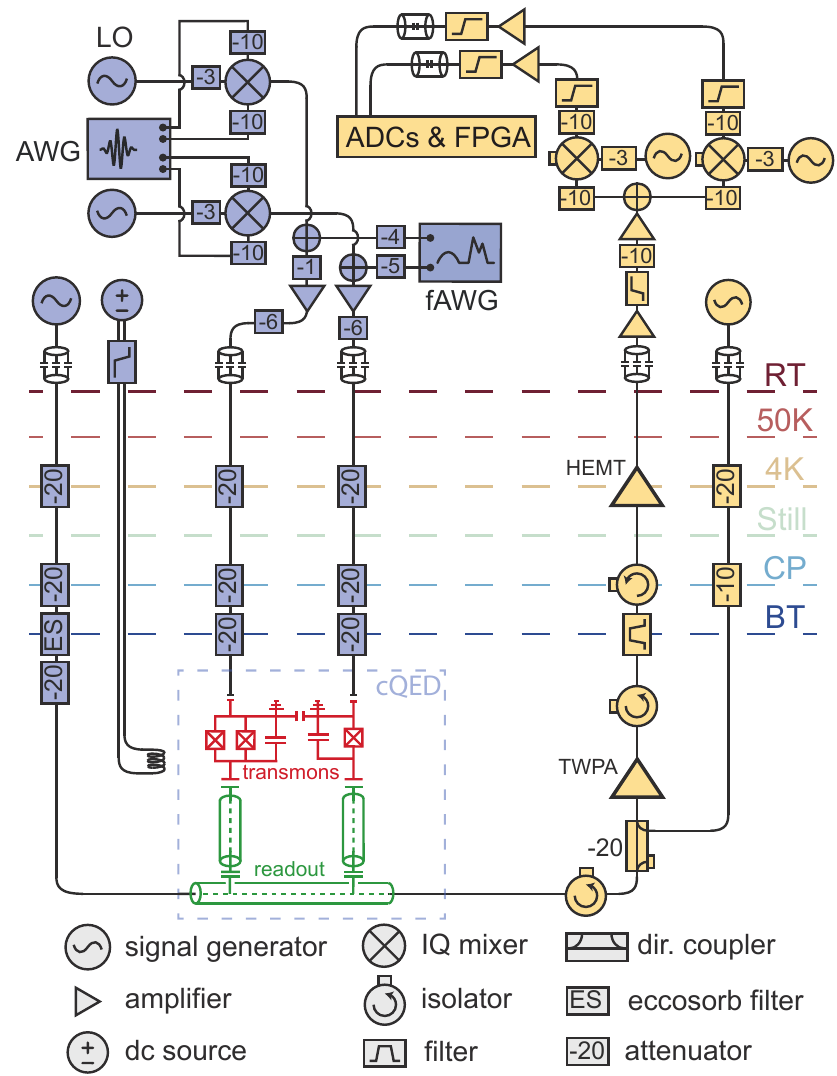}
     \caption{Schematic of the setup. Elements shown in blue indicate input lines, elements shown in yellow indicate detection lines and dashed lines mark the stages of the dilution refrigerator (see text for details).}
     \label{fig7}
\end{figure}

The states of both transmon qubits are read out using a gated microwave tone applied to the input port of a common feed line, see Fig.~\ref{fig7}. The output signal is routed through a circulator and a directional coupler, and amplified at $10\;\mathrm{mK}$ with $24\;\;\mathrm{dB}$ gain at $\omega_{\mathrm{r,A}}/2\pi$ and $20\;\;\mathrm{dB}$ at $\omega_{\mathrm{r,B}}/2\pi$ using a traveling wave parametric amplifier (TWPA), see Fig.~\ref{fig7}. The TWPA is pumped at a frequency of $7.916\;\rm{GHz}$ and we obtain a phase-preserving detection efficiency of $\eta = 0.14$ for the full detection line. The signal is then further amplified by a high-electron-mobility transistor (HEMT) at $4\;\rm{K}$ and two low-noise amplifiers at room temperature. Subsequently, the signal is down-converted to $250\;\rm{MHz}$ using an analog mixer, lowpass-filtered, digitized by an analog-to-digital converter and processed by a field-programmable gate array (FPGA).

We extract the qutrit population of each transmon using single-shot readout with an averaged correct assignment probability of $92\%$ for qubit A and $93\%$ for qubit B. We obtain a maximal correct assignment probability with an integration time of $t_{\rm{m,A}} = 900 \;\rm{ns}$ ($t_{\rm{m,B}} = 960\;\rm{ns}$) and a measurement power that results in a state-dependent photon number in readout resonator A (B) of 102, 2, 86 (28, 18, 29) photons for the states $|g\rangle$, $|e\rangle$, $|f\rangle$, respectively. These photon numbers are below the critical photon numbers \cite{Blais2004} for readout resonator A (B) of 223, 113 (77, 40) for the states $|e\rangle$, $|f\rangle$, respectively.

We extract the parameters of the readout circuit and the relevant coupling strengths from fits to transmission spectrum measurements. The coherence times and anharmonicity of the qutrits are determined in standard time-resolved measurements. All relevant device parameters are listed in Table~\ref{tab:ParameterSummary}.

\begin{table}[b]
\footnotesize
\begin{tabular}{|l|rr|}
\hline
quantity, symbol (unit) & \hfill{}A\hfill{} & \hfill{}B\hfill{} \tabularnewline
\hline
readout resonator frequency, $\omega_{\mathrm{r}}/2\pi$~(GHz)                  & 7.3789  & 7.0762 \tabularnewline
readout resonator bandwidth, $\kappa/2\pi$~(MHz)                       & 0.671   & 0.633   \tabularnewline
readout circuit dispersive shift, $\chi/2\pi$~(MHz)                    & 0.680   & 0.280   \tabularnewline
qubit transition frequency, $\omega_{\mathrm{ge}}/2\pi$~(GHz)                  & 6.4961  & 4.9962  \tabularnewline
transmon anharmonicity, $\alpha/2\pi$~(MHz)                                 &-257.4   &-271.4   \tabularnewline
qubit-qubit coupling strength, $J/2\pi$~(MHz)                                 & \multicolumn{2}{c}{$42\pm 1$}   \tabularnewline
energy relaxation time on $ge$, $T_{\mathrm{1ge}}$~($\mathrm{\mu s}$)  & $7.7\pm0.7$    & $26\pm6$ \tabularnewline
energy relaxation time on $ef$, $T_{\mathrm{1ef}}$~($\mathrm{\mu s}$)  &  $4.4\pm0.5$   & $9\pm2$ \tabularnewline
coherence time on $ge$, $T^{\rm{R}}_{\mathrm{2ge}}$~($\mathrm{\mu s}$) &   $10.3\pm0.6$  & $17\pm4$ \tabularnewline
coherence time on $ef$, $T^{\rm{R}}_{\mathrm{2ef}}$~($\mathrm{\mu s}$) &   $2.7\pm0.5$  & $2.2\pm2$ \tabularnewline
\hline
\end{tabular}
\caption{\label{tab:ParameterSummary} Summary of device parameters for qubit A and qubit B, respectively.}
\end{table}

\section{Calibration of AC-Stark Shift and Coupling Strength}
\label{app:calibration}
To go beyond Eq.~(\ref{eq:fggeLinearCoupling}), which expresses the coupling strength $g_{\rm{fgge}}$ as a linear function of the drive amplitude $\Omega_{\rm{fgge}}$, we numerically diagonalize the system Hamiltonian in dependence on the drive amplitude. For each drive amplitude, we aim to extract both the resonant drive frequency $\omega_{\rm fgge}$ of the fg-ge transition and the coupling strength.
In order to take into account the effect of the drive and the cosine potential of the Josephson junctions fully, we model the coupled two-transmon system in the lab frame,
\begin{equation}
\begin{aligned}
\hat H_{\rm{Floquet}}/\hbar ={}& \sum_{i=A,B} 4 E_{C,i} \hat n_i^2 - E_{J,i} \cos(\hat \varphi_i)\\
& + \tilde J \hat n_A \hat n_B + \Omega(t) \hat n_A,
\end{aligned}
\end{equation}
where all transmon operators are taken in the charge basis, $\tilde J$ is set by the coupling capacitance between the two transmons, and $E_{C,i}$ and $E_{J,i}$ are the charging energy and Josephson energy of transmon $i$, respectively. We consider a drive of the form $\Omega_{\rm A}(t) = \Omega_{\rm fgge} \cos(\omega t)$, and find the resonance frequency $\omega_{\rm{fgge}}(\Omega_{\rm{fgge}})$.

We first set the drive amplitude to zero, $\Omega_{\rm{fgge}}=0$, and choose the parameters $\{E_{C,i},E_{J,i},\tilde J\}$ in order to reproduce the independently extracted parameters listed in Table \ref{tab:ParameterSummary}. We then perform a numerical spectroscopy experiment, and extract $g_{\rm{fgge}}(\Omega_{\rm{fgge}})$ and $\omega_{\rm{fgge}}(\Omega_{\rm{fgge}})$ for each drive amplitude $\Omega_{\rm{fgge}}$ following an approach similar to Ref.~\cite{Zeytinoglu2015}. Essentially, we fix $\Omega_{\rm{fgge}}$ and scan the drive frequency $\omega$, diagonalizing the Hamiltonian for different values of $\omega$. We then extract $g_{\rm{fgge}}$ and $\omega_{\rm{fgge}}$ from the anti-crossing between the states closest to $(|f,g\rangle \pm |g,e\rangle)/\sqrt{2}$.
We extend the numerical protocol in two major ways compared to Ref.~\cite{Zeytinoglu2015}. First, we consider the Hamiltonian in the charge basis, which allows to take into account the full cosine potential of the two transmons. Due to the large drive amplitude, higher states than the $|f\rangle$ state of the transmons are populated. We obtain a maximum population of the $|h\rangle$ ($|i\rangle$) state of 30\% (6\%). Modeling the system Hamiltonian in the charge basis instead of taking an anharmonic oscillator basis allows to describe these states more accurately. Second, we consider the full effect of the drive and find the Floquet eigenmodes \cite{Grifoni1998} of the system in the lab frame instead of performing a rotating-wave approximation (RWA) and diagonalizing a time-independent Hamiltonian in the rotating frame of the drive. This allows to accurately describe the drive since the largest amplitudes considered here correspond to a significant fraction of the drive frequency, $A_{\rm{fgge}} = 1 \rightarrow \Omega_{\rm{fgge}}/\omega_{\rm{fgge}} \approx 0.15$.

To compare the numerical curves with the experimental data, we fit an amplitude conversion factor $A_{\rm{fgge}} = C_{\rm{conv}}\times \Omega_{\rm fgge}$ over the small drive amplitude range $0\leq A_{\rm{fgge}} \leq 0.6$. While Fig.~\ref{fig2}(c) and the inset of Fig.~\ref{fig3}(b) show discrepancies between the data and the numerical model for $A_{\rm{fgge}}>0.7$, Fig.~\ref{fig3}(b) does not depend on the drive amplitude and shows good agreement between the numerical (black line) and experimental data (blue dots) with independently determined parameters. Considering that the simulations agree well with the experimental data when comparing quantities not sensitive to the drive amplitude, we suggest that the discrepancies observed in Fig.~\ref{fig2}(c) and the inset of Fig.~\ref{fig3}(b) are due to the conversion factor between $A_{\rm{fgge}}$ and $\Omega_{\rm fgge}$ depending on frequency, i.e. $C_{\rm{conv}} = C_{\rm{conv}}(\omega_{\rm{fgge}})$.

This can be the case if the drive line of qubit A has a frequency-dependent response or if there are secondary coupling paths from the drive line to the qubit. We verified that the drive line section between the output of the arbitrary waveform generator instrument and the printed circuit board on which the chip is mounted has no frequency dependence beyond the weakly increasing attenuation as a function of frequency characteristic for semi-rigid microwave cables (see e.g. Fig.~13~a in \cite{Krinner2019}), which only explains a $2\%$ deviation of $C_{\rm{conv}}$ at maximum drive amplitude compared to its low amplitude value. However, an impedance mismatch between PCB and the on-chip part of the drive line could introduce a larger frequency dependence. Considering secondary coupling paths, it is possible that in addition to the direct coupling path from drive line to qubit A, a second path is mediated by the readout resonator of qubit A, which has a frequency $\omega_{\mathrm{r,A}}/2\pi = 7.379$\,GHz. The contribution of such a second path is expected to become larger as $\omega_{\rm fgge}$ gets closer to $\omega_{\mathrm{r,A}}$ and the effective $\Omega_{\rm fgge}$ would be given by the interference of both paths.

\section{Calibration of Conditional Phase}
\label{app:CphaseCalibration}
The total conditional phase $\phi_c = \phi_{\rm{fgge}} + \phi_{\rm{zz}}$ accumulated during the gate is a combination of the geometric phase $\phi_{\rm{fgge}}$ and the dynamical phase $\phi_{\rm{zz}} = -\chi t_g$ due to the dispersive coupling. In order to obtain a total phase of $\phi_c = \pi$, the geometric phase should consequently be adjusted to
\begin{equation}
\label{eq:piPhaseCondition}
\phi_{\rm{fgge}} = \pi + \chi t_g.
\end{equation}

This geometric phase can be computed by considering the evolution in the effective $|fg\rangle,|ge\rangle$ two-level system.
Denoting the effective Pauli matrices $\tilde X = |fg\rangle \langle ge | + |ge\rangle \langle fg| $ and $\tilde Z = |fg\rangle \langle fg | - |ge\rangle \langle ge|$, we write an effective two-level Hamiltonian for the driven system

 \begin{equation}
\hat H_{\rm{fgge}} = g_{\rm{fgge}} \tilde X + \frac{1}{2}\Delta_{\rm{fgge}} \tilde Z,
 \end{equation}
 where $\Delta_{\rm{fgge}} = \omega_{\rm{d}} - \omega_{\rm{fgge}}$ is the detuning between the drive and the fgge transition frequency.
After a time $t_g = \pi/\sqrt{g_{\rm{fgge}}^2 + (\Delta_{\rm{fgge}}/2)^2}$, an initial $|ge\rangle$ state completes one round trip in the fg-ge manifold and accumulates a geometric phase $\phi_{\rm{fgge}} = \pi - \Delta_{\rm{fgge}}t_g/2$. From Eq.~(\ref{eq:piPhaseCondition}) we then obtain that the detuning should be set to $\Delta_{\rm{fgge}} = -2 \chi$.

In the experiment, the coupling $g_{\rm{fgge}}$ is not turned on and off instantaneously and, moreover, the dispersive coupling is altered during the gate due to the dressing between the drive and the qubit, $\chi = \chi(\Omega_{\rm{fgge}})$. As a result, the detuning has to be calibrated and we find an optimal working point at $\Delta_{\rm{fgge}}/2\pi = 0.96$ MHz.

\bibliography{Q:/USERS/Sebastian//RefDB/QudevRefDB}
\end{document}